\documentclass[aps,prb,twocolumn,groupedaddress,showpacs]{revtex4}
\usepackage{graphicx}
\usepackage{amsmath}

\newcommand{\eref}[1]{eqn.(\ref{#1})}
\newcommand{\Eref}[1]{Eqn.(\ref{#1})}
\newcommand{\erefs}[2]{eqns.(\ref{#1},\ref{#2})}
\newcommand{\otot}{\hat{1}}
\newcommand{\op}{\hat{1}_{+}}
\newcommand{\om}{\hat{1}_{-}}

\begin{document}

\title{Quantum Phase Transition in Quantum Dot Trimers}

\author{Andrew K. Mitchell, Thomas F. Jarrold, David E. Logan}
\affiliation{Oxford University, Chemistry Department, Physical and Theoretical Chemistry, South Parks Road, Oxford OX1 3QZ, UK}


\date{\today}

\begin{abstract}
We investigate a system of three tunnel-coupled semiconductor quantum dots in a triangular geometry, one 
of which is connected to a metallic lead, in the regime where each dot is essentially singly occupied. 
Both ferro- and antiferromagnetic spin-$\tfrac{1}{2}$ Kondo regimes, separated by a quantum phase transition, 
are shown to arise on tuning the interdot tunnel couplings and should be accessible experimentally. Even in 
the ferromagnetically-coupled local moment phase, the Kondo effect emerges in the vicinity of the transition 
at finite temperatures. Physical arguments and numerical renormalization group techniques are used to 
obtain a detailed understanding of the problem.

\end{abstract}


\maketitle

\section{Introduction}
Nanofabrication techniques developed over the last 
decade~\cite{Goldhaber,scienceartmol,blick2dot,ludwig3dot,vidan3dot,gaudreau3dot,rogge3dot,Fujisawananotube}, together with atomic-scale manipulation using scanning tunneling microscopy~\cite{Crtrimercrommie,Cotrimeruchihashi}, have sparked intense interest in novel mesoscopic devices where strong electron correlation and many-body effects play a central role~\cite{glazman}. The classic spin-$\tfrac{1}{2}$ Kondo effect~\cite{hewson}  -- in which a single spin is screened by antiferromagnetic coupling to conduction electrons in an attached metallic lead -- has been observed in odd-electron semiconductor quantum dots, single molecule dots, and adatoms on metallic surfaces~\cite{glazman}; although its ferromagnetic counterpart, where the spin remains asymptotically free, has yet to be reported experimentally.
In \emph{coupled} 
dot devices -- the `molecular' analogue of single quantum dots viewed as artificial atoms~\cite{scienceartmol} -- exquisite experimental control is now 
available over geometry,
capacitance, and tunnel-couplings of the dots~\cite{vidan3dot,gaudreau3dot}. Both spin and internal, orbital degrees of freedom  -- and hence the interplay between the two -- are important in such systems.
This leads to greater diversity in potentially observable correlated electron behaviour on coupling to metallic leads, as evident from a wide range of  theoretical studies of double 
(e.g. Refs.~\onlinecite{kikoinDQD,vojta2spin,bordaDQD,galpinccdqd,AKMccDQD,andersgalpinnano,lopez,zarandDQD,ingersentDQD}) 
and triple 
(e.g. Refs.~\onlinecite{lazarovitsadatom,hewsonTQD,zitkoTQD,ferrodots,lobosTQD,wangtqd,nutshell,delgadoTQD}) 
quantum dot systems.

Motivated in part by recent experiments involving triple dot devices~\cite{ludwig3dot,vidan3dot,gaudreau3dot,rogge3dot},
we consider here a system of three, mutually tunnel-coupled single-level quantum dots, one of which is 
connected to a metallic lead: a triple quantum dot (TQD) ring structure, the simplest to exhibit frustration.
We focus on the TQD in the 3-electron Coulomb blockade valley, and study its evolution as a function of the 
interdot tunnel couplings, using both perturbative arguments and the full density matrix~\cite{asbasis,fdmnrg} formulation of Wilson's numerical renormalization group (NRG) technique~\cite{nrgreview,KWW} (for a recent review,
see Ref.~\onlinecite{nrgrev}). A rich range of behaviour is found to occur. Both antiferromagnetic  and ferromagnetic 
Kondo physics are shown to arise in the system -- with the two distinct ground states separated by a quantum phase transition -- and should be experimentally accessible via side-gate control of the tunnel couplings.
The zero-bias differential conductance ($G$) through the dots is shown to drop discontinuously across the transition at zero temperature, from the unitarity limit of $G/G_0=1$ in the strong coupling 
antiferromagnetic phase (with $G_0=2e^2/h$ the conductance quantum) to $G/G_0\simeq 0$ in the weak coupling, local 
moment phase. However in a certain temperature window in the vicinity of the transition, the conductance is found to be controlled by the transition fixed point separating the two ground state phases, comprising both a Kondo singlet state and a residual local moment; in particular such that \emph{increasing} temperature in the local moment
phase actually revives the antiferromagnetic Kondo effect.

\begin{figure}[b]
\includegraphics[height=2cm]{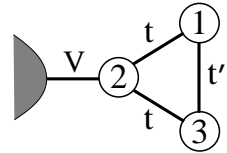}
\caption{\label{dots}Schematic illustration of the quantum dot trimer.}
\end{figure}

\section{Models, bare and effective}
We consider three semiconducting (single-level) quantum dots, arranged in a triangular geometry as illustrated in Fig~\ref{dots}. Each dot is tunnel-coupled to the others, and one of them (dot `2') is also coupled to a metallic lead. We focus explicitly on a system tuned to mirror symmetry (see Fig.~\ref{dots}), and study the Anderson-type model $H=H_0+H_{tri}+H_{hyb}$. Here
$H_0=\sum_{\text{k},\sigma}\epsilon^{\phantom{\dagger}}_{\text{k}} a^{\dagger}_{\text{k} \sigma}a^{\phantom{\dagger}}_{\text{k} \sigma}$ refers to the non-interacting lead, 
which is coupled to dot `2' via
$H_{hyb}=\sum_{\text{k},\sigma} V(a^{\dagger}_{\text{k} \sigma}c^{\phantom{\dagger}}_{2 \sigma}+\text{H.c.})$,
while $H_{tri}$ describes the isolated TQD with tunnel couplings $t,t'$,
\begin{equation}
\label{Htun}
\begin{split}
H_{tri}&=\sum_i(\epsilon \hat{n}_i+U\hat{n}_{i\uparrow}\hat{n}_{i\downarrow}) \\
&~~+\sum_{\sigma} \left[ t~c_{2\sigma}^{\dagger}(c^{\phantom\dagger}_{1\sigma}+
c^{\phantom\dagger}_{3\sigma})
+t'~c_{1\sigma}^{\dagger}c^{\phantom\dagger}_{3\sigma} +\text{H.c.}\right] 
\end{split}
\end{equation}
where $\hat{n}_i=\sum_{\sigma}\hat{n}^{\phantom{\dagger}}_{i\sigma}=\sum_{\sigma}c^{\dagger}_{i \sigma}c^{\phantom{\dagger}}_{i \sigma}$ is the number operator for dot $i$. 
$U$ is the intradot Coulomb repulsion and $\epsilon_i$ the level energy of
dot $i$, such that $\epsilon_{1}=\epsilon_{3} \equiv \epsilon$ for a mirror symmetric system 
(in which $H$ is invariant under a $1\leftrightarrow 3$ permutation). For convenience we take $\epsilon_{i}=\epsilon$ for all dots (although this is not required, as mentioned further below).

We are interested in the TQD deep in the ${\cal{N}}=3$ electron Coulomb blockade valley. To this end, noting that coupled quantum dot experiments typically correspond to $t/U\sim 10^{-2}$~~\cite{gaudreau3dot}, we consider the representative case $\epsilon = -U/2$ with $U\gg (t,t')$. Each dot is then in essence singly occupied, with ${\cal{N}}=2$ or $4$ states much higher ($\sim$$\tfrac{1}{2}U$) in energy. The ${\cal{N}}=3$ 
states of the isolated TQD comprise two lowest doublets, and a spin quartet (which always lies highest in energy).
As the tunnel coupling $t'$ is increased there is a level crossing of the doublets, which are 
degenerate by symmetry at the point $t'=t$. Projected into the singly-occupied manifold the doublet states are
\begin{equation}
\label{eq:states}
\begin{split}
|+; S^{z}\rangle&=c_{2\sigma}^{\dagger}\tfrac{1}{\sqrt{2}}\left(c_{1\uparrow}^{\dagger}
c_{3\downarrow}^{\dagger}+c_{3\uparrow}^{\dagger}c_{1\downarrow}^{\dagger}\right)|\mathrm{vac}\rangle \\
|-;S^{z}\rangle &= \tfrac{\sigma}{\sqrt{6}}\left[ c_{2\sigma}^{\dagger}(c_{1\uparrow}^{\dagger}c_{3\downarrow}^{\dagger}-c_{3\uparrow}^{\dagger}c_{1\downarrow}^{\dagger})-2c_{2-\sigma}^{\dagger}c_{1\sigma}^{\dagger}c_{3\sigma}^{\dagger}\right]
|\mathrm{\mathrm{vac}}\rangle
\end{split}
\end{equation}
with $S^{z}=\tfrac{\sigma}{2}$ and $\sigma =\pm$ for spins $\uparrow/\downarrow$.
Their energy separation is $E_{\Delta}=E_{+}-E_{-}=J-J'$ with antiferromagnetic exchange couplings
$J=4t^2/U$ and $J'=4t'^2/U$, such that the levels cross at $t'=t$ (reflecting the magnetic frustration
inherent at this point). The $|-;S^{z}\rangle$ doublet, containing  triplet configurations of spins `1' and `3',
has odd parity ($-$) under a $1\leftrightarrow 3$ interchange; while $|+;S^{z}\rangle$, which has singlet-locked spins `1' and `3' and behaves in~effect as a spin-$\tfrac{1}{2}$ carried by dot `2' alone, has even parity ($+$).

On coupling to the lead the effective model describing the system on low energy/temperature scales 
is obtained by a standard Schrieffer-Wolff transformation~\cite{SW,hewson}. Provided 
the doublets are not close to degeneracy, only the lower such state need be retained in the ground state manifold:
$|- ;S^{z}\rangle$ for $J'\ll J$ and $|+ ;S^{z}\rangle$ for $J'\gg J$. In either case a low-energy model
of Kondo form arises
\begin{equation}
\label{eq:Jeff}
H_{\text{\emph{eff}}}=J_{K\gamma}~\hat{\textbf{S}}\cdot \hat{\textbf{S}}(0),
\end{equation}
(potential scattering is omitted for clarity),
with $\hat{\textbf{S}}(0)$ the conduction band spin density at dot `2', and $\hat{\textbf{S}}$ a spin-$\tfrac{1}{2}$ operator 
representing the appropriate doublet ($\gamma = +$ or $-$). The effective Kondo
coupling is  $J_{K\gamma}=2\langle\gamma;+\tfrac{1}{2}|\hat{s}^{z}_{2}|\gamma;+\tfrac{1}{2}\rangle J_{K}$ with 
$\hat{\textbf{s}}_{2}$ the spin of dot `2'; where $J_K= 8\Gamma/(\pi\rho U)$ with hybridization
$\Gamma =\pi V^{2}\rho$ and $\rho$ the lead density of states. 
\Eref{eq:states} thus gives $J_{K-}=-\tfrac{1}{3}J_K$, and $J_{K+}=+J_K$.
Hence, for tunnel coupling $t'\ll t$ ($J'\ll J$), a \emph{ferro}magnetic spin-$\tfrac{1}{2}$ Kondo
effect arises ($J_{K-}<0$)~\cite{ferrodots}. Kondo quenching of the lowest doublet is in consequence
ineffective, and as temperature $T\rightarrow 0$ the spin becomes asymptotically free -- the stable fixed point (FP) is the local moment (LM) FP with a residual entropy $S_{\mathrm{imp}}(T=0) =\ln 2$ ($k_{B} =1$)~\cite{KWW}. The system here is the simplest example of  a `singular Fermi liquid'~\cite{mehta}, reflected in the non-analyticity of leading irrelevant corrections to the fixed point~\cite{mehta,singularfm}. For $t'\gg t$ by contrast, the Kondo coupling is antiferromagnetic ($J_{K+}=+J_K>0$), destabilizing the LM fixed point. The strong coupling (SC) FP then controls the $T\rightarrow 0$ behaviour, describing the familiar Fermi liquid Kondo singlet ground state in which the spin is screened by the lead/conduction electrons below the characteristic Kondo scale $T_{K}$, with $T_{K}/\sqrt{U\Gamma} \sim \exp(-1/\rho J_{K+})=\exp(-\pi U/8\Gamma)$~\cite{note2}.

  Since the fixed points for the two stable phases are distinct, a quantum phase transition must thus occur 
on tuning the tunnel coupling $t'$ through a critical value $t'_{c}\simeq t$. We study it below, but first 
outline the effective low-energy model in the vicinity of the transition. Here, as the $|\pm;S^{z}\rangle$ states
are of course near degenerate, both doublets must thus be retained in 
the low-energy trimer manifold, and the unity operator for the local (dot) Hilbert space is hence:
\begin{equation}
\label{eq:unity}
\otot = \sum_{S^{z}} (|+;S^{z}\rangle\langle +;S^{z}| +|-;S^{z}\rangle\langle -;S^{z}|)  
~\equiv ~ \op +\om 
\end{equation}
The effective low-energy model then obtained by Schrieffer-Wolff is readily shown to be
\begin{equation}
\label{eq:fpa}
H_{\text{\emph{eff}}}^{\text{trans}}=J_K ~\otot \hat{\textbf{s}}_{2}^{\phantom\dagger}\otot \cdot \hat{\textbf{S}}(0)+\tfrac{1}{2}E_{\Delta}(\op - \om )
\end{equation}
with $J_{K}$ as above. The final term here refers simply to the energy difference between the two doublets.
It may be written equivalently as $E_{\Delta}\hat{\mathcal{T}}_z$ with a pseudospin operator 
\begin{equation}
\label{eq:pseud}
\hat{\mathcal{T}}_z =\tfrac{1}{2}(\op - \om ) 
\end{equation}
thus defined, such that the doublets are each
eigenstates of it, $\hat{\mathcal{T}}_z |\pm\ ;S^{z} \rangle=\pm\tfrac{1}{2} |\pm ;S^z\rangle$.
Considering now the first term in \eref{eq:fpa},
$\otot \hat{\textbf{s}}_{2}^{\phantom\dagger}\otot \equiv \op \hat{\textbf{s}}_{2}^{\phantom\dagger}\op
+\om \hat{\textbf{s}}_{2}^{\phantom\dagger}\om$ for the mirror symmetric case considered  (cross terms vanish by symmetry). Direct evaluation of $\hat{1}_{\pm} \hat{\textbf{s}}_{2}^{\phantom\dagger}\hat{1}_{\pm}$ gives
$\op \hat{\textbf{s}}_{2}^{\phantom\dagger}\op = \hat{\textbf{S}}\op$ ($=\op\hat{\textbf{S}}$)
and $\om \hat{\textbf{s}}_{2}^{\phantom\dagger}\om = -\tfrac{1}{3}\om\hat{\textbf{S}}$,
where $\hat{\textbf{S}}$ is a spin-$\tfrac{1}{2}$ operator for the dot Hilbert space (specifically
$\hat{S}^{z}= \sum_{\gamma =\pm,~S^{z}}|\gamma;S^{z}\rangle S^{z}\langle\gamma;S^{z}|$ and
$\hat{S}^{\pm}= \sum_{\gamma}|\gamma;\pm\tfrac{1}{2}\rangle\langle\gamma;\mp\tfrac{1}{2}|$).

Hence, using \erefs{eq:unity}{eq:pseud} to express 
$\hat{1}_{\pm}~=~\tfrac{1}{2}(\hat{1} \pm 2 \hat{\mathcal{T}}_z)$ in terms of the pseudospin,
the effective low-energy model is given from \eref{eq:fpa} by
\begin{equation}
\label{eq:fp}
H_{\text{\emph{eff}}}^{\text{trans}}=\tfrac{1}{3} J_K(1+4\hat{\mathcal{T}}_z)\hat{\textbf{S}}\cdot \hat{\textbf{S}}(0)+E_{\Delta}\hat{\mathcal{T}}_z,
\end{equation}
expressed as desired in terms of the spin $\hat{\textbf{S}}$ and pseudospin $\hat{\mathcal{T}}_z$.
The term $E_{\Delta}\hat{\mathcal{T}}_z$ is equivalent to a magnetic field acting on the pseudospin, favoring the $|-;S^{z}\rangle$ doublet for $E_{\Delta}>0$ and $|+;S^{z}\rangle$ for $E_{\Delta}<0$; such that
\eref{eq:fp} reduces, as it should, to one or other of \eref{eq:Jeff} in the limit where the separation 
$|E_{\Delta}|$ is sufficiently large that only one of the doublets need be retained in the low-energy
TQD manifold. Finally, note that the absence of pseudospin raising/lowering terms $\hat{\mathcal{T}}^{\pm}$ in
$H_{\text{\emph{eff}}}^{\text{trans}}$ reflects the strict $1\leftrightarrow 3$ parity in the mirror symmetric setup (which cannot be broken by virtual hopping processes between dot `2' and the lead); and means that the Hilbert space of \eref{eq:fp} separates exactly into spin and pseudospin sectors, such that only the \emph{sign} of the effective Kondo coupling is correlated to the pseudospin.

\section{Results}
The physical picture is thus clear, and indicates the presence of a quantum phase transition as a 
function of $t'$. We now present NRG results for the TQD Anderson model, using a symmetric, constant 
lead density of states $\rho = 1/(2D)$. The full density matrix extension~\cite{asbasis,fdmnrg} of the NRG is employed~\cite{nrgrev}, together with direct calculation of the electron self-energy~\cite{UFG}. Calculations are typically performed for an NRG discretization parameter $\Lambda =3$, retaining the lowest $N_{s}= 3000$ states per iteration.

As above we choose $\epsilon=-\tfrac{1}{2}U$ and $t/U=10^{-2}$, which realistic case~\cite{gaudreau3dot} corresponds to single occupancy of the dots  (all calculations give $\langle \hat{n}_i \rangle=1$ for each dot). The low temperature behaviour is determined by three fixed points: those for the two stable phases at $T=0$ (SC or LM), and a `transition fixed point'  precisely at the transition, which at \emph{finite}-$T$ strongly affects the behaviour of the system close to
the transition.

\begin{figure}[t]
 \includegraphics[height=5.4cm]{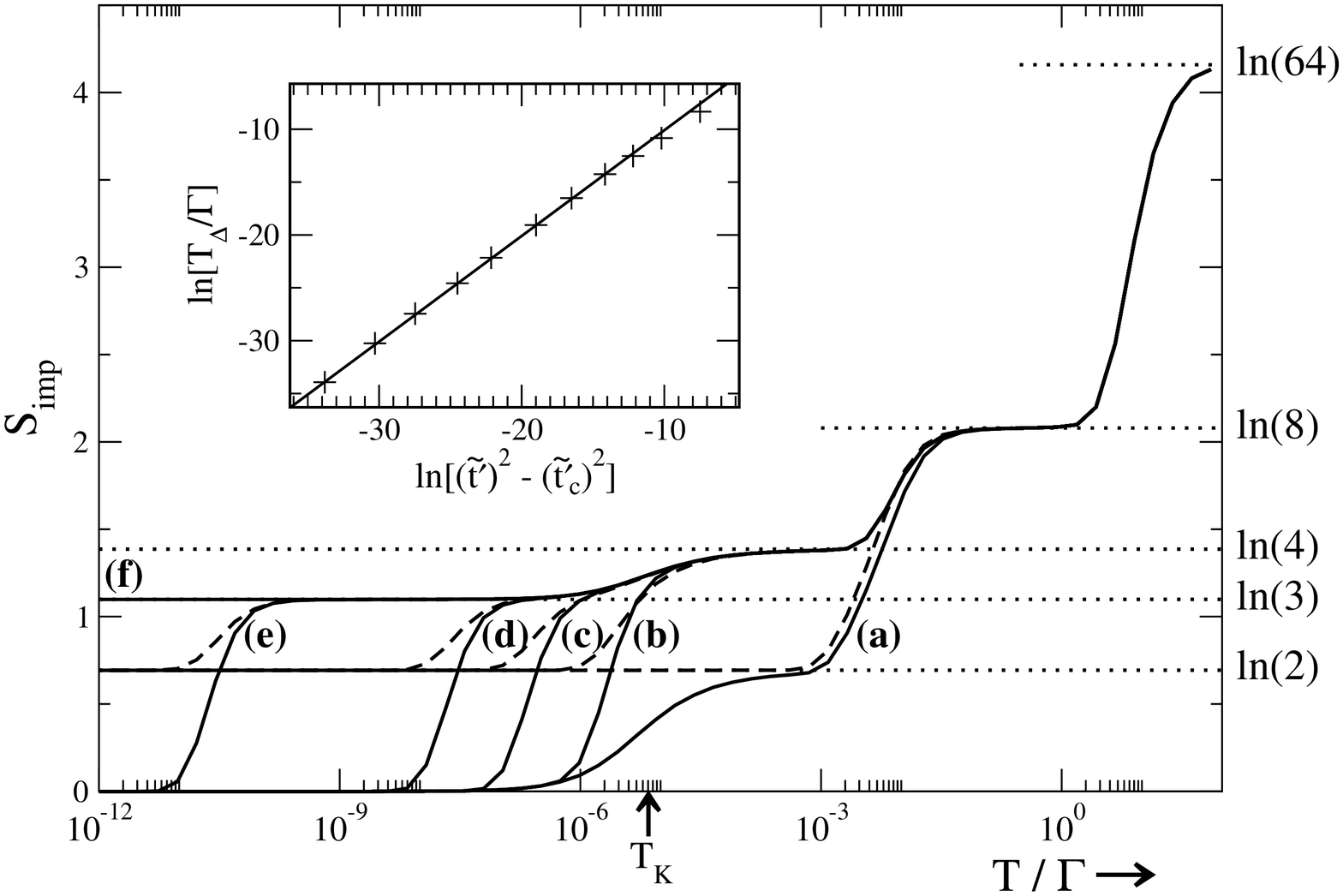}
 \caption{\label{entgraph}
Entropy $S_{\mathrm{imp}}(T)$ \emph{vs} $T/\Gamma$ for fixed $\tilde{U}=10$
and $\tilde{t}=0.1$, as a function of $\tilde{t}'$ as the transition ($\tilde{t}'_{c} \simeq 0.097$)
is approached from either side: for $t'=t'_{c}\pm \lambda T_{K}$ with
$\lambda = 10^4,10,1,10^{-1},10^{-4}$ and $\simeq 0$ (curves (a)-(f) respectively). 
Solid lines: antiferromagnetically coupled SC phase ($\lambda >0$); dashed lines:  
ferromagnetically coupled LM phase ($\lambda <0$).
Inset: close to the transition, the scale $T_{\Delta}$ vanishes linearly in $(\tilde{t'}^2-\tilde{t'_c}^2)$.
}
 \end{figure}

The $T$-dependence of the entropy $S_{\mathrm{imp}}(T)$~\cite{KWW} provides a clear picture
of the relevant fixed points. We show it in Fig.~\ref{entgraph}, for $\tilde{U}=U/\pi\Gamma=10$
and $\tilde{t}=t/\pi\Gamma=0.1$ (with $\Gamma/D = 10^{-2}$), for variable
$\tilde{t'}=t'/\pi\Gamma$ approaching the transition from either side:
$t'=t'_{c}\pm \lambda T_{K}$, varying $\lambda$. Here $\tilde{t}_{c}'=0.09715..(\simeq \tilde{t}$ as expected), and the antiferromagnetic Kondo scale $T_K/\Gamma\simeq 7 \times 10^{-6}$~~\cite{note2}.
Solid lines refer to systems in the SC phase ($t'>t'_{c}$), dashed lines to the LM phase  ($t'<t'_{c}$). 
In all cases the highest $T$ behaviour is governed by the free orbital fixed point~\cite{KWW}, with all $4^3$ states of the TQD thermally accessible and hence $S_{\text{imp}}=\ln (64)$. On the scale $T\sim U$ the dots become singly occupied, the entropy thus dropping to ln($8$).

On further lowering $T$ deep in either the LM or SC phases (case (a) in Fig.~\ref{entgraph}), all but the lowest trimer doublet is projected out and $S_{\text{imp}}$ approaches $\ln(2)$, signifying the LM fixed point.
For $t'<t'_{c}$ this remains the stable fixed point down to $T=0$, while for $t'>t'_c$ the antiferromagnetic
Kondo effect drives the system to the SC fixed point below $T\sim T_K$.
Lines (b)-(e) in Fig.~\ref{entgraph} are for systems progressively approaching the transition.
Here, when $T$ exceeds the energy gap between the doublets (denoted $|\tilde{E}_{\Delta}|$ and naturally renormalized slightly from the isolated TQD limit of $|E_{\Delta}|$), the pair of doublets are effectively degenerate~\cite{note3} and an $S_{\text{imp}} = \ln (4)$ plateau is thus reached. 
The fixed point Hamiltonian here is then simply a free conduction band, plus two free spins (\eref{eq:fp} with $J_K$ and $E_{\Delta}$ set to zero).

\begin{figure}[t]
 \includegraphics[height=5.4cm]{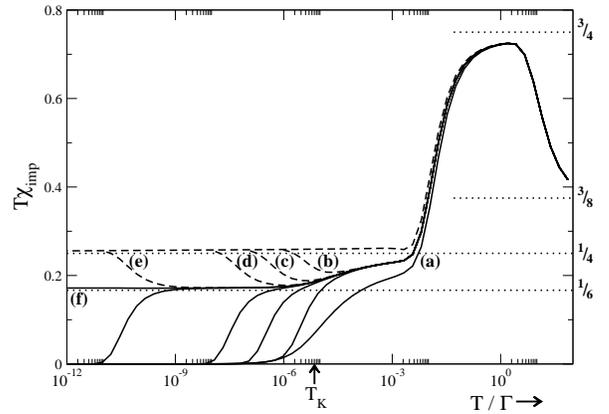}
 \caption{\label{chigraph} Spin susceptibility $T\chi_{\mathrm{imp}}(T)$
\emph{vs} $T/\Gamma$ for the same parameters as Fig.~\ref{entgraph}.
Solid lines: antiferromagnetically coupled SC phase; dashed lines:  
ferromagnetically coupled LM phase. Close to the transition, for $|\tilde{E}_{\Delta}| <T<T_{K}$, a 
$T\chi_{\mathrm{imp}}(T) = \tfrac{1}{6}$ plateau symptomatic of the transition fixed point arises, persisting down to $T=0$ precisely at the transition where $\tilde{E}_{\Delta}=0$ (case (f)).
}
 \end{figure}

For $t'$ within $\sim T_{K}$ of $t'_{c}$, as in (c)-(e) of Fig.~\ref{entgraph}, a further
decrease in $T$ leads to a clear entropy plateau of $S_{\text{imp}} = \ln (3)$. This is
the transition fixed point (TFP): for $|\tilde{E}_{\Delta}| <T<T_{K}$, the $|+;S^{z}\rangle$ doublet is  screened
by the antiferromagnetic Kondo effect -- \emph{even when it is not the ground state} -- 
while the ferromagnetically coupled $|-;S^{z}\rangle$ remains a local moment.
The TFP thus comprises both a free local moment and a Kondo singlet, hence the ln($3$)
entropy. And the TFP Hamiltonian corresponds to \eref{eq:fp} with $E_{\Delta}=0$, 
$J_{K}\rightarrow \infty$ in the $\mathcal{T}_z =+\tfrac{1}{2}$ pseudospin sector
and $J_{K}\rightarrow 0$ for $\mathcal{T}_z =-\tfrac{1}{2}$. The energy level spectrum
at the TFP thus comprises a set of LM levels plus a set of SC levels (as confirmed
directly from the NRG calculations).

Finally, on a scale $T=T_{\Delta} \sim |\tilde{E}_{\Delta}|$, defined in practice
by $S_{\text{imp}}(T_{\Delta}) =0.85$ (suitably between $\ln2$ and $\ln3$), $S_{\text{imp}}$ 
crosses over from the TFP value of $\ln(3)$ to the $T=0$ value appropriate to the stable fixed 
point (SC or LM). As the transition is approached, the scale $T_{\Delta} \propto (t'^2-t'^2_c)$ vanishes  (Fig.~\ref{entgraph} inset) -- a natural consequence of the doublet level crossing, recalling from above that
$T_{\Delta} \approx |E_{\Delta}| = |J'-J|=4|t'^2-t^2|/U$ (with $t'_{c} \equiv t$ for the isolated TQD). And since $T_{\Delta}=0$ precisely at the transition, the $T=0$ entropy at that point is the TFP value $\ln(3)$, as in case (f) of  Fig.~\ref{entgraph}.

The behaviour described for $S_{\mathrm{imp}}(T)$ is likewise evident in the $T$-dependence
of $T\chi_{\mathrm{imp}}^{\phantom\dagger}(T)$ (with $\chi_{\mathrm{imp}}^{\phantom\dagger}(T)$ the
total impurity/dot contribution to the uniform magnetic susceptibility~\cite{KWW}), as shown in 
Fig.~\ref{chigraph}. For $T\gtrsim U$ governed by the free orbital fixed point, 
$T\chi_{\mathrm{imp}}^{\phantom\dagger}(T) \simeq 3\times\tfrac{1}{8}$ as expected~\cite{KWW}
(we set $g\mu_{B} \equiv 1$).
On the scale $T\sim U$ the dots become singly occupied but the spins are essentially uncorrelated,
so $T\chi_{\mathrm{imp}}^{\phantom\dagger}(T) \simeq 3\times \tfrac{1}{4}$ as expected for
three free spins~\cite{KWW}.
On further decreasing $T$, the ultimate low-temperature behaviour is naturally $T\chi_{\mathrm{imp}}^{\phantom\dagger}(T) = \tfrac{1}{4}$ for the ground state doublet characteristic 
of the LM phase $t'<t'_{c}$,
and $T\chi_{\mathrm{imp}}^{\phantom\dagger}(T) = 0$ for the quenched SC fixed point when  $t'>t'_{c}$.
Close to the transition however, for $|\tilde{E}_{\Delta}| <T<T_{K}$, the TFP is again evident
in the persistence of a $T\chi_{\mathrm{imp}}^{\phantom\dagger}(T) = \tfrac{1}{6}$ plateau, 
readily understood as the mean $\langle (S^{z})^{2}\rangle$ for the three quasidegenerate states arising for $|\tilde{E}_{\Delta}| <T<T_{K}$ as described above; and to which value $T\chi_{\mathrm{imp}}^{\phantom\dagger}(T)$ tends as $T \rightarrow 0$, precisely at the transition (Fig.~\ref{chigraph}, case (f)).

\begin{figure}[t]
 \includegraphics[height=5.4cm]{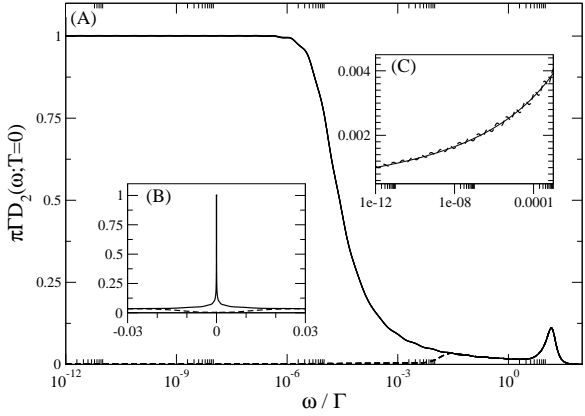}
 \caption{\label{freqdep}$T=0$ local spectrum $\pi\Gamma D_2(\omega;T=0)$ \emph{vs} $\omega /\Gamma$ for systems with the same parameters as Fig.~\ref{entgraph}. Panels (A) and (B) show all data on logarithmic and linear frequency scales respectively, with solid lines again for systems in the SC phase ($t'>t'_c$), and dashed lines for
the LM phase ($t'<t'_c$). Inset (C) shows the low-frequency behaviour in the LM phase, fitted to a line of the form $\pi\Gamma D_2(\omega;T=0)=a/\text{ln}^2(|\omega| / T_0)$.}
 \end{figure}

Most importantly, the physics above is clearly manifest in transport properties. 
The zero-bias conductance through the `2' dot is~\cite{wingreenspec}
\begin{equation}
\label{zbc}
G(T)/G_0=\pi\Gamma \int_{-\infty}^{\infty} -\frac{\partial f(\omega)}{\partial \omega}~ D_2(\omega;T) ~\text{d}\omega ,
\end{equation}
with $f(\omega)$ the Fermi function
and $D_2(\omega;T)$ the local single-particle spectrum
of dot `2'; such that $G(T=0)/G_{0} = \pi\Gamma D_{2}(0;0)$.
Fig.~\ref{freqdep} shows the $T=0$ spectrum for the `2' dot.
Solid (dashed) lines are again for systems in the SC (LM) phase.
At $T=0$, the $\omega =0$ spectral density collapses abruptly as the transition is crossed -- from the unitarity limit of $\pi\Gamma D_2(0;0)=1$ in the SC phase ($t'>t'_c$) to $\pi\Gamma D_2(0;0)=0$ in the LM phase ($t'<t'_c$). All systems in the SC phase share a common Kondo scale, so all solid lines in practice coincide, and are
found to be characterized by the universal scaling form obtained for the single-impurity Anderson model~\cite{costidynamics}. In the LM phase, all $T=0$ spectra again coincide. In this case however,
the low-$\omega$ behaviour is described by $\pi\Gamma D_2(\omega;T=0)\sim a/\text{ln}^2(|\omega| / T_0)$  (as shown in Fig.~\ref{freqdep} inset~(C)), as expected for a `singular Fermi liquid'~\cite{singularfm}, in which the slow approach to the fixed point is characterized by marginally irrelevant logarithmic corrections~\cite{mehta,singularfm}.

\begin{figure}[t]
 \includegraphics[height=5.4cm]{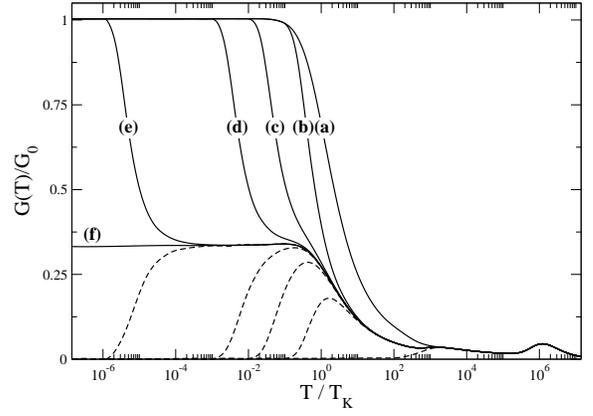}
 \caption{\label{tempdep} Conductance $G(T)/G_0$ \emph{vs} $T/T_{K}$ for the same parameters as Fig.~\ref{entgraph}. As $T\rightarrow 0$, all systems in the SC phase (solid lines) satisfy the unitarity limit $G(T=0)/G_0=1$, while in the LM phase (dashed lines) $G(T=0)/G_0=0$. When $T_{\Delta}<T<T_K$, the conductance is 
controlled by the transition FP, with $G(T)/G_0=\tfrac{1}{3}$ in \emph{both} phases.}
\end{figure}

The zero-bias conductance as a function of $T/T_{K}$ is itself shown in Fig.~\ref{tempdep}.
At $T=0$, $G(0)/G_{0} = 1$ or $0$ in the SC or LM phases respectively, as above.
Far from the transition (case (a)), $G(T)$ decays steadily with increasing $T$ in the SC phase, reflecting thermal destruction of the coherent Kondo singlet. In the LM phase by contrast, $G(T)/G_{0}$ is not appreciable at any $T$. Close to the transition however, and for $T_{\Delta}\lesssim T\lesssim T_K$, the transition FP controls the zero-bias conductance, and $G(T)/G_0=\tfrac{1}{3}$ is seen in both SC \emph{and} LM phases -- meaning in particular that \emph{warming} a system in the ferromagnetically coupled LM phase produces a revival of the
\emph{antiferromagnetic} Kondo effect. This behaviour is readily understood by noting that at $T=0$ (as in Fig.~\ref{freqdep}), $D_{2}(\omega; 0) \equiv D_{2}^{LM}(\omega)$ for $t'<t'_{c}$, as only excitations from the doubly degenerate LM ground states are relevant; while for $t'>t'_{c}$, $D_{2}(\omega;0) \equiv D_{2}^{SC}(\omega)$ since now only excitations from the Kondo singlet arise. At finite-$T$ however, with $T_{\Delta}\ll T \ll T_K$ such that the lowest manifold of states comprises both the LM and the Kondo singlet, it is easy to show from the Lehmann representation of the spectrum that $D_{2}(\omega;T) = \tfrac{1}{3}[D_{2}^{SC}(\omega)+2D_{2}^{LM}(\omega)]$ (such that $\pi\Gamma D_{2}(\omega;T) = 1/3$ for $|\omega|\lesssim T_{K}$); and hence from Eq.~\ref{zbc} that 
$G(T)/G_{0} = 1/3$ -- which persists down to $T=0$ precisely at the transition, where $T_{\Delta}=0$ (Fig.~\ref{tempdep}, case(f)).

\section{Conclusion}
The TQD ring system in the 3-electron Coulomb blockade valley exhibits a rich range of physical behaviour. Both antiferromagnetic and ferromagnetic spin-$\tfrac{1}{2}$ Kondo physics is accessible
on tuning the tunnel coupling $t'$; the two phases being separated by a level crossing quantum phase transition, reflected in a transition fixed point which controls in particular the conductance in the vicinity of the transition. We also add that while explicit results have been given for a mirror symmetric TQD with $\epsilon_{i}=\epsilon$ ($=-U/2$) for all dots, our conclusions are robust provided the dots remain in essence singly occupied. Varying $\epsilon$ (or $U$) on dot `2' for example, making it inequivalent to dots `1' and `3', does not break mirror symmetry and leaves unaltered the behaviour uncovered above. Indeed even breaking mirror symmetry, via e.g.\ distinct tunnel couplings between all dots, still results in both ferromagnetically-coupled and antiferromagnetically-coupled (Kondo quenched) ground states separated by a quantum phase transition~\cite{note4}. The robustness of the essential physics  suggests that both phases should be experimentally accessible in a TQD device; as too should the transition between them, provided the tunnel couplings can be sufficiently finely
tuned.

\acknowledgements
We are grateful to M. Galpin, C. Wright and T. Kuzmenko for stimulating discussions. This research was supported in part by EPSRC Grant EP/D050952/1.


\begin{thebibliography}{43}
\expandafter\ifx\csname natexlab\endcsname\relax\def\natexlab#1{#1}\fi
\expandafter\ifx\csname bibnamefont\endcsname\relax
  \def\bibnamefont#1{#1}\fi
\expandafter\ifx\csname bibfnamefont\endcsname\relax
  \def\bibfnamefont#1{#1}\fi
\expandafter\ifx\csname citenamefont\endcsname\relax
  \def\citenamefont#1{#1}\fi
\expandafter\ifx\csname url\endcsname\relax
  \def\url#1{\texttt{#1}}\fi
\expandafter\ifx\csname urlprefix\endcsname\relax\def\urlprefix{URL }\fi
\providecommand{\bibinfo}[2]{#2}
\providecommand{\eprint}[2][]{\url{#2}}

\bibitem[{\citenamefont{Goldhaber-Gordon
  et~al.}(1998)\citenamefont{Goldhaber-Gordon, Shtrikman, Mahalu,
  Abusch-Magder, Meirav, and Kastner}}]{Goldhaber}
\bibinfo{author}{\bibfnamefont{D.}~\bibnamefont{Goldhaber-Gordon}},
  \bibinfo{author}{\bibfnamefont{H.}~\bibnamefont{Shtrikman}},
  \bibinfo{author}{\bibfnamefont{D.}~\bibnamefont{Mahalu}},
  \bibinfo{author}{\bibfnamefont{D.}~\bibnamefont{Abusch-Magder}},
  \bibinfo{author}{\bibfnamefont{U.}~\bibnamefont{Meirav}}, \bibnamefont{and}
  \bibinfo{author}{\bibfnamefont{M.~A.} \bibnamefont{Kastner}},
  \bibinfo{journal}{Nature} \textbf{\bibinfo{volume}{391}},
  \bibinfo{pages}{156} (\bibinfo{year}{1998}).

\bibitem[{\citenamefont{Jeong et~al.}(2001)\citenamefont{Jeong, Chang, and
  Melloch}}]{scienceartmol}
\bibinfo{author}{\bibfnamefont{H.}~\bibnamefont{Jeong}},
  \bibinfo{author}{\bibfnamefont{A.~M.} \bibnamefont{Chang}}, \bibnamefont{and}
  \bibinfo{author}{\bibfnamefont{M.~R.} \bibnamefont{Melloch}},
  \bibinfo{journal}{Science} \textbf{\bibinfo{volume}{293}},
  \bibinfo{pages}{2221} (\bibinfo{year}{2001}).

\bibitem[{\citenamefont{Blick et~al.}(1996)\citenamefont{Blick, Haug, Weis,
  Pfannkuche, Klitzing, and Eberl}}]{blick2dot}
\bibinfo{author}{\bibfnamefont{R.~H.} \bibnamefont{Blick}},
  \bibinfo{author}{\bibfnamefont{R.~J.} \bibnamefont{Haug}},
  \bibinfo{author}{\bibfnamefont{J.}~\bibnamefont{Weis}},
  \bibinfo{author}{\bibfnamefont{D.}~\bibnamefont{Pfannkuche}},
  \bibinfo{author}{\bibfnamefont{K.~v.} \bibnamefont{Klitzing}},
  \bibnamefont{and} \bibinfo{author}{\bibfnamefont{K.}~\bibnamefont{Eberl}},
  \bibinfo{journal}{Phys. Rev. B} \textbf{\bibinfo{volume}{53}},
  \bibinfo{pages}{7899} (\bibinfo{year}{1996}).

\bibitem[{\citenamefont{Schr\"{o}er et~al.}(2007)\citenamefont{Schr\"{o}er,
  Greentree, Gaudreau, Eberl, Hollenberg, Kotthaus, and Ludwig}}]{ludwig3dot}
\bibinfo{author}{\bibfnamefont{D.}~\bibnamefont{Schr\"{o}er}},
  \bibinfo{author}{\bibfnamefont{A.~D.} \bibnamefont{Greentree}},
  \bibinfo{author}{\bibfnamefont{L.}~\bibnamefont{Gaudreau}},
  \bibinfo{author}{\bibfnamefont{K.}~\bibnamefont{Eberl}},
  \bibinfo{author}{\bibfnamefont{L.~C.~L.} \bibnamefont{Hollenberg}},
  \bibinfo{author}{\bibfnamefont{J.~P.} \bibnamefont{Kotthaus}},
  \bibnamefont{and} \bibinfo{author}{\bibfnamefont{S.}~\bibnamefont{Ludwig}},
  \bibinfo{journal}{Phys. Rev. B} \textbf{\bibinfo{volume}{76}},
  \bibinfo{eid}{075306} (\bibinfo{year}{2007}).

\bibitem[{\citenamefont{Vidan et~al.}(2005)\citenamefont{Vidan, Westervelt,
  Stopa, Hanson, and Gossard}}]{vidan3dot}
\bibinfo{author}{\bibfnamefont{A.}~\bibnamefont{Vidan}},
  \bibinfo{author}{\bibfnamefont{R.}~\bibnamefont{Westervelt}},
  \bibinfo{author}{\bibfnamefont{M.}~\bibnamefont{Stopa}},
  \bibinfo{author}{\bibfnamefont{M.}~\bibnamefont{Hanson}}, \bibnamefont{and}
  \bibinfo{author}{\bibfnamefont{A.}~\bibnamefont{Gossard}},
  \bibinfo{journal}{J. Supercond. Incorp. Novel Magn.}
  \textbf{\bibinfo{volume}{18}}, \bibinfo{pages}{223} (\bibinfo{year}{2005}).

\bibitem[{\citenamefont{Gaudreau et~al.}(2006)}]{gaudreau3dot}
\bibinfo{author}{\bibfnamefont{L.}~\bibnamefont{Gaudreau}}
  \bibnamefont{et~al.}, \bibinfo{journal}{Phys. Rev. Lett.}
  \textbf{\bibinfo{volume}{97}}, \bibinfo{eid}{036807} (\bibinfo{year}{2006}).

\bibitem[{\citenamefont{Rogge and Haug}(2008)}]{rogge3dot}
\bibinfo{author}{\bibfnamefont{M.~C.} \bibnamefont{Rogge}} \bibnamefont{and}
  \bibinfo{author}{\bibfnamefont{R.~J.} \bibnamefont{Haug}},
  \bibinfo{journal}{Phys. Rev. B} \textbf{\bibinfo{volume}{77}},
  \bibinfo{eid}{193306} (\bibinfo{year}{2008}).

\bibitem[{\citenamefont{Grove-Rasmussen
  et~al.}(2008)\citenamefont{Grove-Rasmussen, J{\o}rgensen, Hayashi, Lindelof,
  and Fujisawa}}]{Fujisawananotube}
\bibinfo{author}{\bibfnamefont{K.}~\bibnamefont{Grove-Rasmussen}},
  \bibinfo{author}{\bibfnamefont{H.~I.} \bibnamefont{J{\o}rgensen}},
  \bibinfo{author}{\bibfnamefont{T.}~\bibnamefont{Hayashi}},
  \bibinfo{author}{\bibfnamefont{P.~E.} \bibnamefont{Lindelof}},
  \bibnamefont{and} \bibinfo{author}{\bibfnamefont{T.}~\bibnamefont{Fujisawa}},
  \bibinfo{journal}{Nano Letters} \textbf{\bibinfo{volume}{8}},
  \bibinfo{pages}{1055} (\bibinfo{year}{2008}).

\bibitem[{\citenamefont{Jamneala et~al.}(2001)\citenamefont{Jamneala, Madhavan,
  and Crommie}}]{Crtrimercrommie}
\bibinfo{author}{\bibfnamefont{T.}~\bibnamefont{Jamneala}},
  \bibinfo{author}{\bibfnamefont{V.}~\bibnamefont{Madhavan}}, \bibnamefont{and}
  \bibinfo{author}{\bibfnamefont{M.~F.} \bibnamefont{Crommie}},
  \bibinfo{journal}{Phys. Rev. Lett.} \textbf{\bibinfo{volume}{87}},
  \bibinfo{pages}{256804} (\bibinfo{year}{2001}).

\bibitem[{\citenamefont{Uchihashi et~al.}(2008)\citenamefont{Uchihashi, Zhang,
  Kr\"{o}ger, and Berndt}}]{Cotrimeruchihashi}
\bibinfo{author}{\bibfnamefont{T.}~\bibnamefont{Uchihashi}},
  \bibinfo{author}{\bibfnamefont{J.}~\bibnamefont{Zhang}},
  \bibinfo{author}{\bibfnamefont{J.}~\bibnamefont{Kr\"{o}ger}},
  \bibnamefont{and} \bibinfo{author}{\bibfnamefont{R.}~\bibnamefont{Berndt}},
  \bibinfo{journal}{Phys. Rev. B} \textbf{\bibinfo{volume}{78}},
  \bibinfo{eid}{033402} (\bibinfo{year}{2008}).

\bibitem[{\citenamefont{Kouwenhoven and Glazman}(2001)}]{glazman}
\bibinfo{author}{\bibfnamefont{L.~P.} \bibnamefont{Kouwenhoven}}
  \bibnamefont{and} \bibinfo{author}{\bibfnamefont{L.~I.}
  \bibnamefont{Glazman}}, \bibinfo{journal}{Physics World}
  \textbf{\bibinfo{volume}{14}}, \bibinfo{pages}{33} (\bibinfo{year}{2001}).

\bibitem[{\citenamefont{Hewson}(1993)}]{hewson}
\bibinfo{author}{\bibfnamefont{A.~C.} \bibnamefont{Hewson}},
  \emph{\bibinfo{title}{The {K}ondo Problem to Heavy Fermions}}
  (\bibinfo{publisher}{Cambridge University Press},
  \bibinfo{address}{Cambridge}, \bibinfo{year}{1993}).

\bibitem[{\citenamefont{Kikoin and Avishai}(2001)}]{kikoinDQD}
\bibinfo{author}{\bibfnamefont{K.}~\bibnamefont{Kikoin}} \bibnamefont{and}
  \bibinfo{author}{\bibfnamefont{Y.}~\bibnamefont{Avishai}},
  \bibinfo{journal}{Phys. Rev. Lett.} \textbf{\bibinfo{volume}{86}},
  \bibinfo{pages}{2090} (\bibinfo{year}{2001}).

\bibitem[{\citenamefont{Vojta et~al.}(2002)\citenamefont{Vojta, Bulla, and
  Hofstetter}}]{vojta2spin}
\bibinfo{author}{\bibfnamefont{M.}~\bibnamefont{Vojta}},
  \bibinfo{author}{\bibfnamefont{R.}~\bibnamefont{Bulla}}, \bibnamefont{and}
  \bibinfo{author}{\bibfnamefont{W.}~\bibnamefont{Hofstetter}},
  \bibinfo{journal}{Phys. Rev. B} \textbf{\bibinfo{volume}{65}},
  \bibinfo{pages}{140405(R)} (\bibinfo{year}{2002}).

\bibitem[{\citenamefont{Borda et~al.}(2003)\citenamefont{Borda, Zar\'and,
  Hofstetter, Halperin, and von Delft}}]{bordaDQD}
\bibinfo{author}{\bibfnamefont{L.}~\bibnamefont{Borda}},
  \bibinfo{author}{\bibfnamefont{G.}~\bibnamefont{Zar\'and}},
  \bibinfo{author}{\bibfnamefont{W.}~\bibnamefont{Hofstetter}},
  \bibinfo{author}{\bibfnamefont{B.~I.} \bibnamefont{Halperin}},
  \bibnamefont{and} \bibinfo{author}{\bibfnamefont{J.}~\bibnamefont{von
  Delft}}, \bibinfo{journal}{Phys. Rev. Lett.} \textbf{\bibinfo{volume}{90}},
  \bibinfo{pages}{026602} (\bibinfo{year}{2003}).

\bibitem[{\citenamefont{Galpin et~al.}(2005)\citenamefont{Galpin, Logan, and
  Krishnamurthy}}]{galpinccdqd}
\bibinfo{author}{\bibfnamefont{M.~R.} \bibnamefont{Galpin}},
  \bibinfo{author}{\bibfnamefont{D.~E.} \bibnamefont{Logan}}, \bibnamefont{and}
  \bibinfo{author}{\bibfnamefont{H.~R.} \bibnamefont{Krishnamurthy}},
  \bibinfo{journal}{Phys. Rev. Lett.} \textbf{\bibinfo{volume}{94}},
  \bibinfo{pages}{186406} (\bibinfo{year}{2005}).

\bibitem[{\citenamefont{Mitchell et~al.}(2006)\citenamefont{Mitchell, Galpin,
  and Logan}}]{AKMccDQD}
\bibinfo{author}{\bibfnamefont{A.~K.} \bibnamefont{Mitchell}},
  \bibinfo{author}{\bibfnamefont{M.~R.} \bibnamefont{Galpin}},
  \bibnamefont{and} \bibinfo{author}{\bibfnamefont{D.~E.} \bibnamefont{Logan}},
  \bibinfo{journal}{Europhys. Lett.} \textbf{\bibinfo{volume}{76}},
  \bibinfo{pages}{95} (\bibinfo{year}{2006}).

\bibitem[{\citenamefont{Anders et~al.}(2008)\citenamefont{Anders, Logan,
  Galpin, and Finkelstein}}]{andersgalpinnano}
\bibinfo{author}{\bibfnamefont{F.~B.} \bibnamefont{Anders}},
  \bibinfo{author}{\bibfnamefont{D.~E.} \bibnamefont{Logan}},
  \bibinfo{author}{\bibfnamefont{M.~R.} \bibnamefont{Galpin}},
  \bibnamefont{and}
  \bibinfo{author}{\bibfnamefont{G.}~\bibnamefont{Finkelstein}},
  \bibinfo{journal}{Phys. Rev. Lett.} \textbf{\bibinfo{volume}{100}},
  \bibinfo{eid}{086809} (\bibinfo{year}{2008}).

\bibitem[{\citenamefont{L\'opez et~al.}(2005)\citenamefont{L\'opez, S\'anchez,
  Lee, Choi, Simon, and Le~Hur}}]{lopez}
\bibinfo{author}{\bibfnamefont{R.}~\bibnamefont{L\'opez}},
  \bibinfo{author}{\bibfnamefont{D.}~\bibnamefont{S\'anchez}},
  \bibinfo{author}{\bibfnamefont{M.}~\bibnamefont{Lee}},
  \bibinfo{author}{\bibfnamefont{M.-S.} \bibnamefont{Choi}},
  \bibinfo{author}{\bibfnamefont{P.}~\bibnamefont{Simon}}, \bibnamefont{and}
  \bibinfo{author}{\bibfnamefont{K.}~\bibnamefont{Le~Hur}},
  \bibinfo{journal}{Phys. Rev. B} \textbf{\bibinfo{volume}{71}},
  \bibinfo{pages}{115312} (\bibinfo{year}{2005}).

\bibitem[{\citenamefont{Zar\'{a}nd et~al.}(2006)\citenamefont{Zar\'{a}nd,
  Chung, Simon, and Vojta}}]{zarandDQD}
\bibinfo{author}{\bibfnamefont{G.}~\bibnamefont{Zar\'{a}nd}},
  \bibinfo{author}{\bibfnamefont{C.-H.} \bibnamefont{Chung}},
  \bibinfo{author}{\bibfnamefont{P.}~\bibnamefont{Simon}}, \bibnamefont{and}
  \bibinfo{author}{\bibfnamefont{M.}~\bibnamefont{Vojta}},
  \bibinfo{journal}{Phys. Rev. Lett.} \textbf{\bibinfo{volume}{97}},
  \bibinfo{eid}{166802} (\bibinfo{year}{2006}).

\bibitem[{\citenamefont{{Dias da Silva} et~al.}(2008)\citenamefont{{Dias da
  Silva}, Ingersent, Sandler, and Ulloa}}]{ingersentDQD}
\bibinfo{author}{\bibfnamefont{L.~G. G.~V.} \bibnamefont{{Dias da Silva}}},
  \bibinfo{author}{\bibfnamefont{K.}~\bibnamefont{Ingersent}},
  \bibinfo{author}{\bibfnamefont{N.}~\bibnamefont{Sandler}}, \bibnamefont{and}
  \bibinfo{author}{\bibfnamefont{S.~E.} \bibnamefont{Ulloa}},
  \bibinfo{journal}{Phys. Rev. B} \textbf{\bibinfo{volume}{78}},
  \bibinfo{eid}{153304} (\bibinfo{year}{2008}).

\bibitem[{\citenamefont{Lazarovits et~al.}(2005)\citenamefont{Lazarovits,
  Simon, Zar\'{a}nd, and Szunyogh}}]{lazarovitsadatom}
\bibinfo{author}{\bibfnamefont{B.}~\bibnamefont{Lazarovits}},
  \bibinfo{author}{\bibfnamefont{P.}~\bibnamefont{Simon}},
  \bibinfo{author}{\bibfnamefont{G.}~\bibnamefont{Zar\'{a}nd}},
  \bibnamefont{and} \bibinfo{author}{\bibfnamefont{L.}~\bibnamefont{Szunyogh}},
  \bibinfo{journal}{Phys. Rev. Lett.} \textbf{\bibinfo{volume}{95}},
  \bibinfo{eid}{077202} (\bibinfo{year}{2005}).

\bibitem[{\citenamefont{Oguri et~al.}(2005)\citenamefont{Oguri, Nisikawa, and
  Hewson}}]{hewsonTQD}
\bibinfo{author}{\bibfnamefont{A.}~\bibnamefont{Oguri}},
  \bibinfo{author}{\bibfnamefont{Y.}~\bibnamefont{Nisikawa}}, \bibnamefont{and}
  \bibinfo{author}{\bibfnamefont{A.~C.} \bibnamefont{Hewson}},
  \bibinfo{journal}{J. Phys. Soc. Jpn.} \textbf{\bibinfo{volume}{74}},
  \bibinfo{pages}{2554} (\bibinfo{year}{2005}).

\bibitem[{\citenamefont{\v{Z}itko et~al.}(2006)\citenamefont{\v{Z}itko,
  Bon\v{c}a, Ram\v{s}ak, and Rejec}}]{zitkoTQD}
\bibinfo{author}{\bibfnamefont{R.}~\bibnamefont{\v{Z}itko}},
  \bibinfo{author}{\bibfnamefont{J.}~\bibnamefont{Bon\v{c}a}},
  \bibinfo{author}{\bibfnamefont{A.}~\bibnamefont{Ram\v{s}ak}},
  \bibnamefont{and} \bibinfo{author}{\bibfnamefont{T.}~\bibnamefont{Rejec}},
  \bibinfo{journal}{Phys. Rev. B} \textbf{\bibinfo{volume}{73}},
  \bibinfo{eid}{153307} (\bibinfo{year}{2006}).

\bibitem[{\citenamefont{Kuzmenko et~al.}(2006)\citenamefont{Kuzmenko, Kikoin,
  and Avishai}}]{ferrodots}
\bibinfo{author}{\bibfnamefont{T.}~\bibnamefont{Kuzmenko}},
  \bibinfo{author}{\bibfnamefont{K.}~\bibnamefont{Kikoin}}, \bibnamefont{and}
  \bibinfo{author}{\bibfnamefont{Y.}~\bibnamefont{Avishai}},
  \bibinfo{journal}{Phys. Rev. B} \textbf{\bibinfo{volume}{73}},
  \bibinfo{eid}{235310} (\bibinfo{year}{2006}).

\bibitem[{\citenamefont{Lobos and Aligia}(2006)}]{lobosTQD}
\bibinfo{author}{\bibfnamefont{A.~M.} \bibnamefont{Lobos}} \bibnamefont{and}
  \bibinfo{author}{\bibfnamefont{A.~A.} \bibnamefont{Aligia}},
  \bibinfo{journal}{Phys. Rev. B} \textbf{\bibinfo{volume}{74}},
  \bibinfo{eid}{165417} (\bibinfo{year}{2006}).

\bibitem[{\citenamefont{Wang}(2007)}]{wangtqd}
\bibinfo{author}{\bibfnamefont{W.~Z.} \bibnamefont{Wang}},
  \bibinfo{journal}{Phys. Rev. B} \textbf{\bibinfo{volume}{76}},
  \bibinfo{eid}{115114} (\bibinfo{year}{2007}).

\bibitem[{\citenamefont{Ferrero et~al.}(2007)\citenamefont{Ferrero, Leo,
  Lecheminant, and Fabrizio}}]{nutshell}
\bibinfo{author}{\bibfnamefont{M.}~\bibnamefont{Ferrero}},
  \bibinfo{author}{\bibfnamefont{L.~D.} \bibnamefont{Leo}},
  \bibinfo{author}{\bibfnamefont{P.}~\bibnamefont{Lecheminant}},
  \bibnamefont{and} \bibinfo{author}{\bibfnamefont{M.}~\bibnamefont{Fabrizio}},
  \bibinfo{journal}{J. Phys.: Condens. Matter} \textbf{\bibinfo{volume}{19}},
  \bibinfo{pages}{433201} (\bibinfo{year}{2007}).

\bibitem[{\citenamefont{Delgado and Hawrylak}(2008)}]{delgadoTQD}
\bibinfo{author}{\bibfnamefont{F.}~\bibnamefont{Delgado}} \bibnamefont{and}
  \bibinfo{author}{\bibfnamefont{P.}~\bibnamefont{Hawrylak}},
  \bibinfo{journal}{J. Phys.: Condens. Matter} \textbf{\bibinfo{volume}{20}},
  \bibinfo{pages}{315207} (\bibinfo{year}{2008}).

\bibitem[{\citenamefont{Peters et~al.}(2006)\citenamefont{Peters, Pruschke, and
  Anders}}]{asbasis}
\bibinfo{author}{\bibfnamefont{R.}~\bibnamefont{Peters}},
  \bibinfo{author}{\bibfnamefont{T.}~\bibnamefont{Pruschke}}, \bibnamefont{and}
  \bibinfo{author}{\bibfnamefont{F.~B.} \bibnamefont{Anders}},
  \bibinfo{journal}{Phys. Rev. B} \textbf{\bibinfo{volume}{74}},
  \bibinfo{eid}{245114} (\bibinfo{year}{2006}).

\bibitem[{\citenamefont{Weichselbaum and von Delft}(2007)}]{fdmnrg}
\bibinfo{author}{\bibfnamefont{A.}~\bibnamefont{Weichselbaum}}
  \bibnamefont{and} \bibinfo{author}{\bibfnamefont{J.}~\bibnamefont{von
  Delft}}, \bibinfo{journal}{Phys. Rev. Lett.} \textbf{\bibinfo{volume}{99}},
  \bibinfo{eid}{076402} (\bibinfo{year}{2007}).

\bibitem[{\citenamefont{Wilson}(1975)}]{nrgreview}
\bibinfo{author}{\bibfnamefont{K.~G.} \bibnamefont{Wilson}},
  \bibinfo{journal}{Rev. Mod. Phys.} \textbf{\bibinfo{volume}{47}},
  \bibinfo{pages}{773} (\bibinfo{year}{1975}).

\bibitem[{\citenamefont{Krishnamurthy et~al.}(1980)\citenamefont{Krishnamurthy,
  Wilkins, and Wilson}}]{KWW}
\bibinfo{author}{\bibfnamefont{H.~R.} \bibnamefont{Krishnamurthy}},
  \bibinfo{author}{\bibfnamefont{J.~W.} \bibnamefont{Wilkins}},
  \bibnamefont{and} \bibinfo{author}{\bibfnamefont{K.~G.}
  \bibnamefont{Wilson}}, \bibinfo{journal}{Phys. Rev. B}
  \textbf{\bibinfo{volume}{21}}, \bibinfo{pages}{1003, 1044}
  (\bibinfo{year}{1980}).

\bibitem[{\citenamefont{Bulla et~al.}(2008)\citenamefont{Bulla, Costi, and
  Pruschke}}]{nrgrev}
\bibinfo{author}{\bibfnamefont{R.}~\bibnamefont{Bulla}},
  \bibinfo{author}{\bibfnamefont{T.}~\bibnamefont{Costi}}, \bibnamefont{and}
  \bibinfo{author}{\bibfnamefont{T.}~\bibnamefont{Pruschke}},
  \bibinfo{journal}{Rev. Mod. Phys.} \textbf{\bibinfo{volume}{80}},
  \bibinfo{pages}{395} (\bibinfo{year}{2008}).

\bibitem[{\citenamefont{Schrieffer and Wolff}(1966)}]{SW}
\bibinfo{author}{\bibfnamefont{J.}~\bibnamefont{Schrieffer}} \bibnamefont{and}
  \bibinfo{author}{\bibfnamefont{P.}~\bibnamefont{Wolff}},
  \bibinfo{journal}{Phys. Rev.} \textbf{\bibinfo{volume}{149}},
  \bibinfo{pages}{491} (\bibinfo{year}{1966}).

\bibitem[{\citenamefont{Mehta et~al.}(2005)\citenamefont{Mehta, Andrei,
  Coleman, Borda, and Zar\'{a}nd}}]{mehta}
\bibinfo{author}{\bibfnamefont{P.}~\bibnamefont{Mehta}},
  \bibinfo{author}{\bibfnamefont{N.}~\bibnamefont{Andrei}},
  \bibinfo{author}{\bibfnamefont{P.}~\bibnamefont{Coleman}},
  \bibinfo{author}{\bibfnamefont{L.}~\bibnamefont{Borda}}, \bibnamefont{and}
  \bibinfo{author}{\bibfnamefont{G.}~\bibnamefont{Zar\'{a}nd}},
  \bibinfo{journal}{Phys. Rev. B} \textbf{\bibinfo{volume}{72}},
  \bibinfo{eid}{014430} (\bibinfo{year}{2005}).

\bibitem[{\citenamefont{Koller et~al.}(2005)\citenamefont{Koller, Hewson, and
  Meyer}}]{singularfm}
\bibinfo{author}{\bibfnamefont{W.}~\bibnamefont{Koller}},
  \bibinfo{author}{\bibfnamefont{A.~C.} \bibnamefont{Hewson}},
  \bibnamefont{and} \bibinfo{author}{\bibfnamefont{D.}~\bibnamefont{Meyer}},
  \bibinfo{journal}{Phys. Rev. B} \textbf{\bibinfo{volume}{72}},
  \bibinfo{eid}{045117} (\bibinfo{year}{2005}).

\bibitem[{not({\natexlab{a}})}]{note2}
\bibinfo{note}{The $U$-dependence of $T_{K}$ extracted from the numerics
  confirms the mapping of the full tunneling $H$ onto a spin-$\tfrac{1}{2}$
  Kondo model, with $T_{K} \propto U\sqrt{\rho J_{K+}} \exp(-1/\rho
  J_{K+})$~\cite{KWW} and $\rho J_{K+} = 8\Gamma/(\pi U)$ as obtained
  perturbatively above}.

\bibitem[{\citenamefont{Bulla et~al.}(1998)\citenamefont{Bulla, Hewson, and
  Pruschke}}]{UFG}
\bibinfo{author}{\bibfnamefont{R.}~\bibnamefont{Bulla}},
  \bibinfo{author}{\bibfnamefont{A.~C.} \bibnamefont{Hewson}},
  \bibnamefont{and} \bibinfo{author}{\bibfnamefont{T.}~\bibnamefont{Pruschke}},
  \bibinfo{journal}{J. Phys.: Condens. Matter} \textbf{\bibinfo{volume}{10}},
  \bibinfo{pages}{8365} (\bibinfo{year}{1998}).

\bibitem[{not({\natexlab{b}})}]{note3}
\bibinfo{note}{The trimer quartet state is always higher in energy, and
  thermally inaccessible for $T \ll \mathrm{max}(4t^{2}/U,4t'^{2}/U)$}.

\bibitem[{\citenamefont{Meir and Wingreen}(1992)}]{wingreenspec}
\bibinfo{author}{\bibfnamefont{Y.}~\bibnamefont{Meir}} \bibnamefont{and}
  \bibinfo{author}{\bibfnamefont{N.~S.} \bibnamefont{Wingreen}},
  \bibinfo{journal}{Phys. Rev. Lett.} \textbf{\bibinfo{volume}{68}},
  \bibinfo{pages}{2512} (\bibinfo{year}{1992}).

\bibitem[{\citenamefont{Costi et~al.}(1994)\citenamefont{Costi, Hewson, and
  Zlati\'{c}}}]{costidynamics}
\bibinfo{author}{\bibfnamefont{T.~A.} \bibnamefont{Costi}},
  \bibinfo{author}{\bibfnamefont{A.~C.} \bibnamefont{Hewson}},
  \bibnamefont{and}
  \bibinfo{author}{\bibfnamefont{V.}~\bibnamefont{Zlati\'{c}}},
  \bibinfo{journal}{J. Phys.: Condens. Matter} \textbf{\bibinfo{volume}{6}},
  \bibinfo{pages}{2519} (\bibinfo{year}{1994}).

\bibitem[{not({\natexlab{c}})}]{note4}
\bibinfo{note}{Here the spin-$\tfrac{1}{2}$ pseudospin is not of course
  conserved, and the transition itself is no longer a simple level crossing.}

\end{thebibliography}

\end{document}